# «EPECUR»[ϒ]

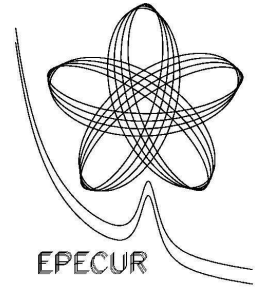

(*Search for the Cryptoexotic Member $N_{\overline{10}}$ of the Baryon Antidecuplet ½⁺ in the Reactions $\bar{\pi}p \rightarrow \bar{\pi}p$ and $\bar{\pi}p \rightarrow K\Lambda$*)

Experiment Proposal from ITEP-PNPI Collaboration.

I.G. Alekseev[*,1], P.Ye. Budkovsky, V.P. Kanavets, M.M. Kats, L.I. Koroleva,
V.V. Kulikov, B.V. Morozov, V.M. Nesterov, V.V. Ryltsov, V.A. Sakharov,
A.D. Sulimov, D.N. Svirida

Institute for Theoretical and Experimental Physics,
B. Cheremushkinskaya 25, 117218, Moscow, Russia

A.I. Kovalev, N.G. Kozlenko, V.S. Kozlov, A.G. Krivshich, D.V. Novinsky,
V.V. Sumachev[*,2], V.Yu. Trautman, Ye.A. Filimonov

Petersburg Nuclear Physics Institute,
Gatchina, Leningrad district 188300, Russia.

*May 2004, revised May 2005*

## Abstract

The main goal of this proposal is the search for a narrow cryptoexotic nucleon resonance by scanning of the $\pi^-p$ system invariant mass in the region (1610–1770) MeV with the detection of $\pi^-p$ and $K\Lambda$ decays. The scan is supposed to be done by the variation of the incident $\pi^-$ momentum and its measurement with the accuracy of up to ±0.1% (better than 1 MeV in terms of the invariant mass in the whole energy range) with a set of proportional chambers located in first focus of the magneto-optical channel. High sensitivity of the method to the resonance under search is shown. The secondary particles scattered from a liquid hydrogen target are detected by sets of the wire drift chambers equipped with modern electronics. The time scale of the project is about 3 years. The budget estimate including manpower, the apparatus and operation cost, is about 40 million rubles. The beam time required is (4–6) two week runs on "high" (10 GeV/c) flattop of the ITEP proton synchrotron.

---

[ϒ] This is a direct transcription of the Russian abbreviation for "Experiment for the Pentaquark Search in the Elastic Scattering"
[*] spokenpersons
[1] E-mail: igor.alekseev@itep.ru
[2] E-mail: sumachev@pnpi.spb.ru

## Introduction

Dramatic events in baryon spectroscopy took place during the last year. The narrow exotic baryon $\theta^+$ with strangeness +1 and mass 1540 MeV (Fig.1) was discovered by LEPS [1] and DIANA at ITEP [2], which was earlier predicted by the chiral soliton model [2]. The results from LEPS and DIANA were confirmed by several successive measurements [5-12]. Due to its quantum numbers, $\theta^+$ can contain four quark and one antiquark as a minimum, and this is why this particle was called pentaquark. In the chiral soliton model $\theta^+$ belongs to $SU(3)_F$ baryon antidecuplet with the spin and parity equal $\frac{1}{2}^+$. This antidecuplet should also contain the cryptoexotic baryons with the quantum numbers of the nucleon and $\Sigma$-hyperon (Fig. 1). No doubts that the discovery of these missing members of the antidecuplet would be a great step towards the understanding of the strong interaction dynamics in the nonperturbative region.

The aim of this proposal is to perform the experimental search for the cryptoexotic non-strange neutral resonance $N_{\overline{10}}$ (isospin projection −1/2) in the reactions $\pi^-p \to \pi^-p$ and/or $\pi^-p \to K\Lambda$. According to the spin/parity of the antidecuplet ($\frac{1}{2}^+$) the resonant effect should be searched in $P_{11}$-wave.

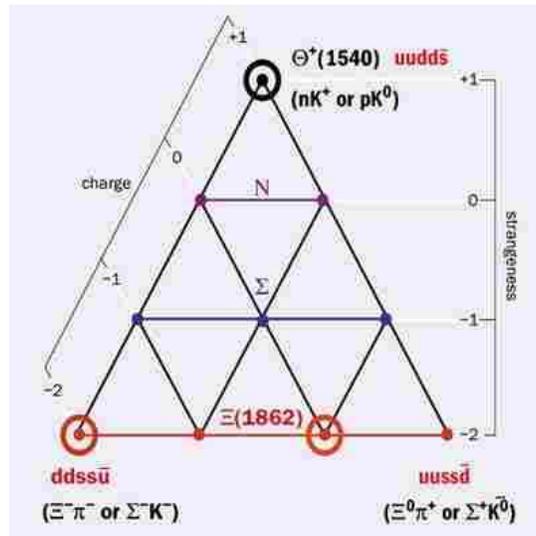

Figure 1. "Pentaquark" antidecuplet

## 1. Experiment Motivation and Layout

The general idea of the proposal is to look for the $N_{\overline{10}}$ effects in the cross-section in the "formation" type experiments using the elastic pion-nucleon scattering and reaction $\pi^-p \to K\Lambda$. The scan of the mass interval under investigation will be done by changing the initial pion momentum. Secondary pion beams with appropriate intensity and energy are available at ITEP from its 10 GeV proton synchrotron. Two-focused beamline optics provides the possibility to analyse the individual pion momentum with the accuracy up to (0.06-0.15)%, having the total momentum range of $\Delta p/p = \pm 2\%$. Wide energy range of (0.8–2.5) GeV/c can be covered by changing of the magnetic elements currents.

The results of the measurements may be analyzed by the standard procedure of the partial-wave (PWA) analysis, which is the important advantage of the "formation" type experiments. In particular it means that all the quantum numbers of the resonance, if found, can be unambiguously determined.

## 1.1. Background information

The existing data on non-exotic $P_{11}$ resonance in 1700 MeV region, the theoretical estimates of $N_{\overline{10}}$ parameters and estimates of the expected effect are presented in the following section.

### 1.1.1. "Common" resonance N(1710)

The following summary on the non-exotic $P_{11}$ resonance N(1710) is taken mainly from the "Review of Particle Properties" (RPP, [13]). This resonance has a rating of three stars (***). According to the results of the energy independent partial wave analyses (IPWA): KH80 [14] – $M_R$=1723±9 MeV, Γ=120±15 MeV; CMB80 [15] – $M_R$=1700±50 MeV, Γ=90±30 MeV. The multi-channel analysis of the πN→Nπ and πN→Nππ partial waves by KSU [16] has two solutions, predicting either a wide resonance with Γ=480±230 MeV or a narrow one with Γ=50±40 MeV at the same mass of $M_R$=1717±28 MeV. In the recent energy dependent (DPWA) analysis FA02 [17] of GWU group the resonance N(1710) is not observed. The estimates for the branching ratios for Nπ and KΛ channels are (10-20)% and (5-25)% respectively. In spite of rather high PDG rating, the resonance parameters are poorly defined and even its existence is doubtful, suggesting further experimental studies in corresponding energy region.

### 1.1.2. "Pentaquarks" $N_{\overline{10}}$ and θ⁺.

The summary of experimental results on the mass and width of θ⁺ is presented in the following table:

| Collaboration | Mass $M_R$, MeV | Width Γ, MeV | Reference |
|---|---|---|---|
| LEPS | 1540±10 | <25 | [1] |
| DIANA | 1539±2 | <9 | [2] |
| CLAS/γd | 1542±5 | <21 | [5] |
| SAPHIR | 1540±6 | <25 | [6] |
| ITEP | 1533±5 | <20 | [7] |
| CLAS/γp | 1555±10 | <26 | [8] |
| HERMES | 1526±3 | 13±9 | [9] |
| ZEUS | 1522±3 | 8±4 | [10] |
| COSY-TOF | 1530±5 | <18 | [11] |
| SVD | 1526±5 | <24 | [12] |

In case of the negligible mixing with the non-exotic baryon octet it is naturally to expect the total width Γ($N_{\overline{10}}$) to be similar to the width of θ⁺. Careful study of the possible properties of $N_{\overline{10}}$ was made by the authors of [18] both from the theoretical (chiral soliton model) and phenomenological ("modified" PWA) points of view. Their detailed investigation of the "modified" FA02 analysis near 1700 MeV showed indications of $N_{\overline{10}}$ in two regions having the mass 1680 MeV (more probably) and/or 1730 MeV, while the theoretical estimate based on the σ-term value from this PWA leads to the prediction of 1650 MeV. The resonance elastic width was estimated as $Γ_{EL}$($N_{\overline{10}}$)≤0.5 MeV. The authors of [18] also note that in the framework of the standard PWA procedure based on the existing experimental data resonance with a width Γ<30 MeV may be easily lost. The case of mixing with the baryon octet [N(1440), Λ(1600), Σ(1660), Ξ(1690)] was analyzed in [19]. These estimates lead to $N_{\overline{10}}$ mass in the interval (1650-1690) MeV. The most probable width of $N_{\overline{10}}$ is ~10 MeV

(does not exceed 30 MeV), partial widths of the decay to Nπ and KΛ are ~2MeV and ~1MeV respectively.

Latest experimental indications for $N_{\overline{10}}$ are not too convincing. Yet one should mention:
- The results from STAR [20] where a candidate was claimed to be seen at 1733.6±5 MeV with the width upper estimate of 6 MeV in $K^0_s\Lambda$ from Au-Au collisions at 200 GeV.
- The resonance-like structure at 1675 MeV seen by GRAAL in the η–photoproduction on neutron differential cross-section [21]. The same collaboration got preliminary indication of a resonance peak near 1730 MeV in γn→$K^0\Lambda$ and γn→$K^+\Sigma^-$ [22]

### 1.1.3. Observable Choice -- Some Formulas

The formulas presented in this section are intended to illustrate the choice of the observables that should be measured in π⁻p interactions to get the optimal sensitivity for a resonant effect.

The decomposition of the scattering amplitude into partial waves:

$$f_{el}(\theta, E) = \frac{1}{2ik}\sum_{l=0}^{l_{max}}(2l+1)(\eta_l \cdot e^{2i\delta_l} - 1)P_l(\cos\theta)$$

where $\delta_l$ is the phase, $\eta_l$ – the elasticity of l-wave and the term

$$f_l = \frac{\eta_l \cdot e^{2i\delta_l} - 1}{2ik}$$

is called partial amplitude. In the vicinity of a (narrow) resonance the partial amplitude can be written in the form:

$$f_l(E) = f_l^B + f_l^r(E)$$

where $f_l^B$ – the "background" term, weakly dependent on the energy, while $f_l^r(E)$ is the resonant term, dominantly determining the local energy dependence. The corresponding partial cross-section will be:

$$\sigma_l = 4\pi(2l+1)|f_l^B + f_l^r|^2 \qquad (1)$$

The following considerations will be presented in parallel for the both reactions mentioned in this proposal in order to get direct comparison of the two cases.

| π⁻p elastic | π⁻p→KΛ |
|---|---|
| The Breit-Wigner resonant amplitude: | |
| $f_l^r(E) = -\dfrac{e^{i\varphi}\Gamma_{el}}{2k[(E-E_r)+i\Gamma/2]}$ | $f_l^r(E) = -\dfrac{e^{i\varphi}\sqrt{\Gamma_{el}\Gamma_{K\Lambda}}}{2\sqrt{kk_{K\Lambda}}[(E-E_r)+i\Gamma/2]}$ |
| The BW amplitude value in the resonance | |
| $f_l^r(E_r) = i\dfrac{e^{i\varphi}X}{k}$, | $f_l^r(E_r) = i\dfrac{e^{i\varphi}\sqrt{X \cdot BR}}{\sqrt{kk_{K\Lambda}}}$, |
| where $X = \dfrac{\Gamma_{el}}{\Gamma}$ is the branching ratio to the π⁻p channel | where $BR = \dfrac{\Gamma_{K\Lambda}}{\Gamma}$ is the branching ratio to the KΛ channel |

Neglecting the interference term in (1) for a rough estimate, the resonant effect in the total reaction cross-section will be:

$$\Delta\sigma_{el}^r = (2l+1)\cdot\frac{4\pi}{k^2}\cdot X^2 \qquad (2a) \qquad\qquad \Delta\sigma_{K\Lambda}^r = (2l+1)\cdot\frac{4\pi}{kk_{K\Lambda}}\cdot X \cdot BR \qquad (2b)$$

For $P_{11}$ wave ($l=1$), taking the resonance mass $m_r=1.7$ GeV, $k=0.56$ GeV/c, $k_{K\Lambda}=0.2$ GeV/c (both at $W_{cm}=m_r$), and assuming $X=0.01$, $BR=0.1$:

$$\Delta\sigma^r_{el}=0.005 \text{ mb} \qquad | \qquad \Delta\sigma^r_{K\Lambda}=0.13 \text{ mb}$$

Or in terms of the relative cross-section change ($\sigma_{el}\approx 10$ mb, $\sigma_{K\Lambda}\approx 0.9$ mb)

$$\Delta\sigma^r_{el}/\sigma_{el} = 0.05\% \qquad | \qquad \Delta\sigma^r_{K\Lambda}/\sigma_{K\Lambda} = 15\%$$

The conclusion is obvious: the total reaction cross-section of $\pi^-p\to K\Lambda$ is highly sensitive to the resonant effect, even assuming very conservative values for the branching ratios. At the same time the total elastic cross-section has very low sensitivity to resonances with low $X$, because of the $X^2$ nature of the effect. Instead, as will be shown in the next section, the differential cross-section of this process can be used for the search, because it reveals and, in certain angular regions, emphasizes the contribution of the interference terms containing the products of the resonant wave amplitude and the amplitudes of other (non-resonant) waves. It's worth mentioning that this contribution is proportional to $X$, and not $X^2$.

### 1.1.4. Expected effects

In case of the elastic scattering it is possible to accurately evaluate the sensitivity of the differential cross-section to the resonant change of the phase $\delta_{P11}$ and of the elasticity $\eta_{P11}$ of $P_{11}$ wave using the PWA results. Fig. 2 shows the relative change of $\pi^-p$ elastic differential cross-section as a function of c.m. scattering angle, when the $\eta_l$ parameter of various waves is changed by a certain amount, normalized by this amount:

$$S_l(\theta_{CM}) = \frac{1}{\left(d\sigma/d\Omega\right)} \cdot \frac{\partial\left(d\sigma/d\Omega\right)}{\partial\eta_l}$$

These dependencies were obtained from FA02 analysis [17] for the initial pion momentum 1000 MeV/c ($W_{cm}=1671$ MeV). One can see that for the $P_{11}$-wave (thick red curve) the high sensitivity interval lays from 40° to 90°. This interval corresponds to the cross-section minimum with $d\sigma/d\Omega\sim 0.2$ mb/sr (Fig. 3). The maximum sensitivity is 1.5 per unit change of $\eta_{P11}$ and is reached at approximately 65° c.m. scattering angle. The sensitivity to the change of $\delta_{P11}$ phase is small in the whole region under consideration.

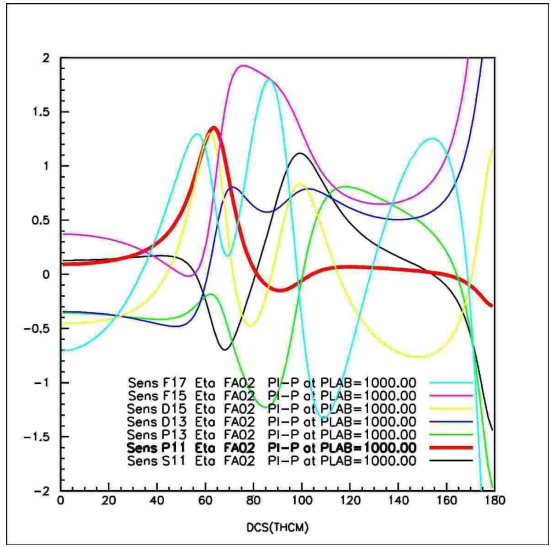
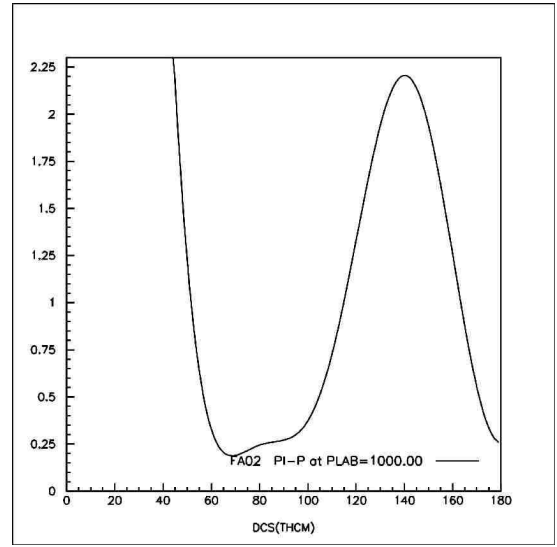

Figure 2. Relative sensitivity $S_l$ of the $\pi^-p$ elastic differential cross-section at 1000 MeV/c for various partial waves.

Figure 3. $\pi^-p$ elastic differential cross-section at 1000 MeV/c from FA02 [17].

Now let us consider an Argand diagram for a narrow resonance on top of a slowly changing non-resonant amplitude in a certain partial wave (fig. 4). The length of the vector from the center of the unitary circle to the "root" of the resonance represents the non-resonant amplitude elasticity $\eta_B/2$, while the diameter of the resonance ring is $X=\Gamma_{el}/\Gamma$. Thus in the approximation $X << \eta_B \sim 1$ the change of the amplitude elasticity $\eta_l$ due to the resonance is $\Delta\eta_l \approx 2\cdot\Gamma_{el}/\Gamma$. In the same approximation the resonant effect in the phase is $\Delta\delta_l \approx (\Gamma_{el}/\Gamma)/\eta_l$. Similar conclusion can be drawn from the formalism in [18], where it is shown that the amplitude change due to the resonance is close to $r_a=X=\Gamma_{el}/\Gamma$.

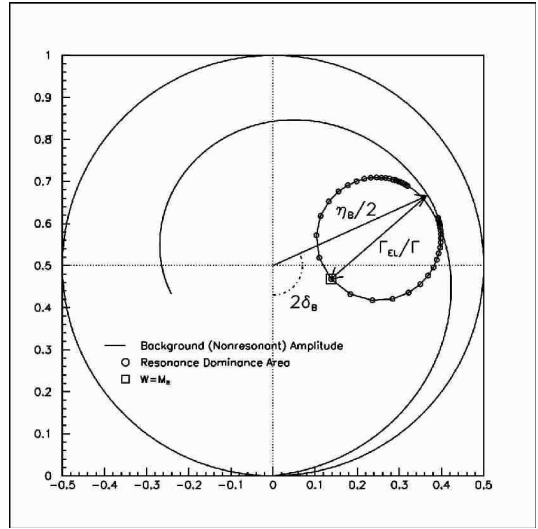

Figure 4. Argand diagram for a narrow resonance in a certain partial wave. For the purpose of better visibility the resonance is shown with large $X=\Gamma_{el}/\Gamma$ compared to that expected for $N_{\overline{10}}$.

Since the resonant effect in the differential cross-section can be estimated as $\Delta\eta_l\cdot S_l$ and supposing for $N_{\overline{10}}$, for example, $X(N_{\overline{10}})=5\%$, one gets 15% deviation from the non-resonant behavior in the differential cross-section in the area of sensitivity maximum. Such high sensitivity have a simple explanation: in the differential cross-section minimum all partial waves compensate each other, each of them having large but sign alternating contribution; in this case even a small disturbance in one of the waves will result in a significant relative change of the cross-section.

Fig. 5 illustrates how $N_{\overline{10}}$ could be seen in the elastic scattering assuming its full width equal 6 MeV, elasticity $X=5\%$ and mass 1671 MeV. The insertion in the top right corner is the zoom of the area around the resonance. Error bars and the point density correspond to the proposed statistical accuracy and momentum resolution. It's worth mentioning that the effect of the resonance can be either a minimum (as in the figure), a maximum or a bipolar structure, dependent on its unpredictable pole residue phase.

As shown above, the reaction $\pi^-p \to K\Lambda$ is even more promising for the search of the resonances with small elasticity. Compared to the elastic case the cross-section of the process itself is low, making its resonant change relatively more pronounceable. The sensitivity of this process to the resonant effects is significantly enhanced by the fact that the energy region under study is close to the reaction threshold (mind the term $k_{K\Lambda}$ in the denominator of (2b)). Another argument in favor of this reaction is that the branching ratio of $N_{\overline{10}}$ to this channel is predicted larger than that to $\pi^-p$ [18].

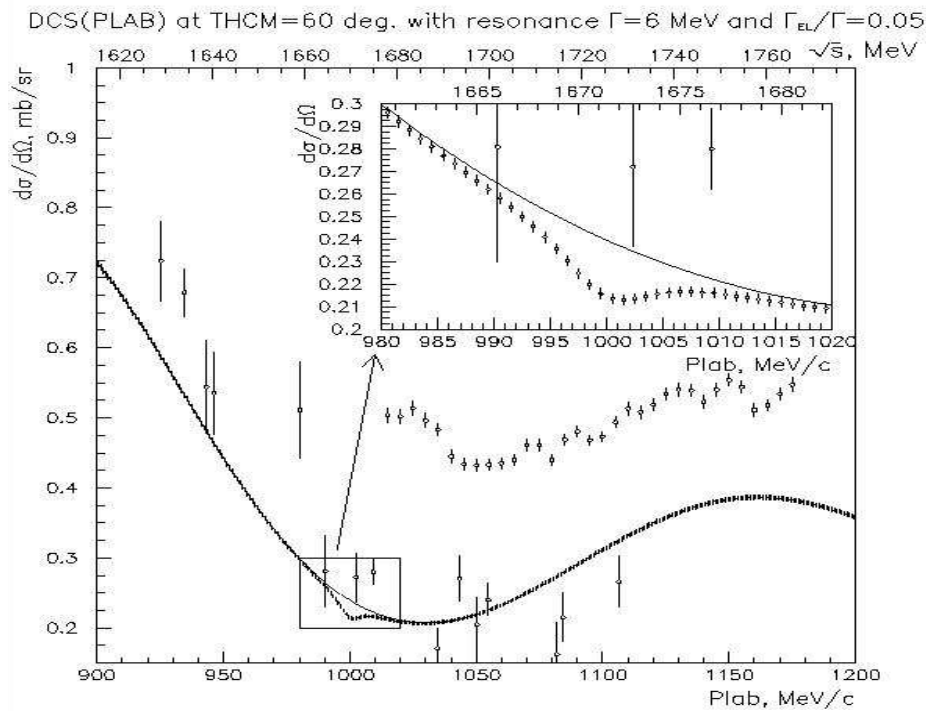

Figure 5. $\pi^-$p elastic differential cross-section at 60° c.m. as a function of pion momentum in the presence of $N_{\overline{10}}$; the upper scale is the c.m. energy; the most accurate of the existing data are at a different angle.

The process $\pi^-p \to K\Lambda$ has an important additional advantage because of its "selfanalyzing" feature. Due to the parity violation in week $\Lambda$-decays, the polarization of this particle can be extracted from the angular distribution of the reaction products. Thus the two observables, cross-section and normal polarization, are measured simultaneously. The peculiarity in the energy behavior of the polarization also can indicate the presence of a resonance. From the other side the new data on this parameter may significantly improve the PWA of $\pi^-p \to K\Lambda$ process.

### *1.2. General Requirements to the Setup*

The large uncertainties in the expected parameters of the resonance under search lead to the following requirements to setup:
- The coverage of mass interval (1610-1770) MeV
- The mass resolution ≤1MeV
- The statistics must be sufficient for the registration of resonance with elasticity ≤5% and branching ratio to $K\Lambda$ channel ≤10%
- High detector efficiency
- Long term efficiency stability (with precision ~1%) to get undistorted energy dependence in the mass interval ~160 MeV
- The possibility of the internal crosscheck of the results
- The reliable background suppression
- High accuracy of track detectors (better than 0.15 mm) for K and $\Lambda$ decay vertex reconstruction
- Minimum amount of matter on the way of initial particles and reaction products (momentum resolution and vertex reconstruction)

## 1.3. Setup for $N_{\overline{10}}$ Resonance Search in the Elastic Scattering

### 1.3.1. Setup Layout

The layout of the setup for $N_{\overline{10}}$ search in $\pi^-p \to \pi^-p$ elastic scattering is shown in Fig. 6. The setup consists of the following main elements:

- Proportional chambers PC1–PC3 located in first focus of the ITEP beamline 322 for the incident pion momentum measurement
- Proportional chambers PC4–PC6 for the tracking of the incident pions
- Liquid hydrogen target with 1 mm beryllium cover, 4 cm in diameter, 25 cm long
- Sets of drift chambers DC1–DC4 for the reaction products tracking
- Beam scintillation counters S1, S2, trigger hodoscopes H2,H3 with time of flight measurements features, anti- and beam TOF counter A1.

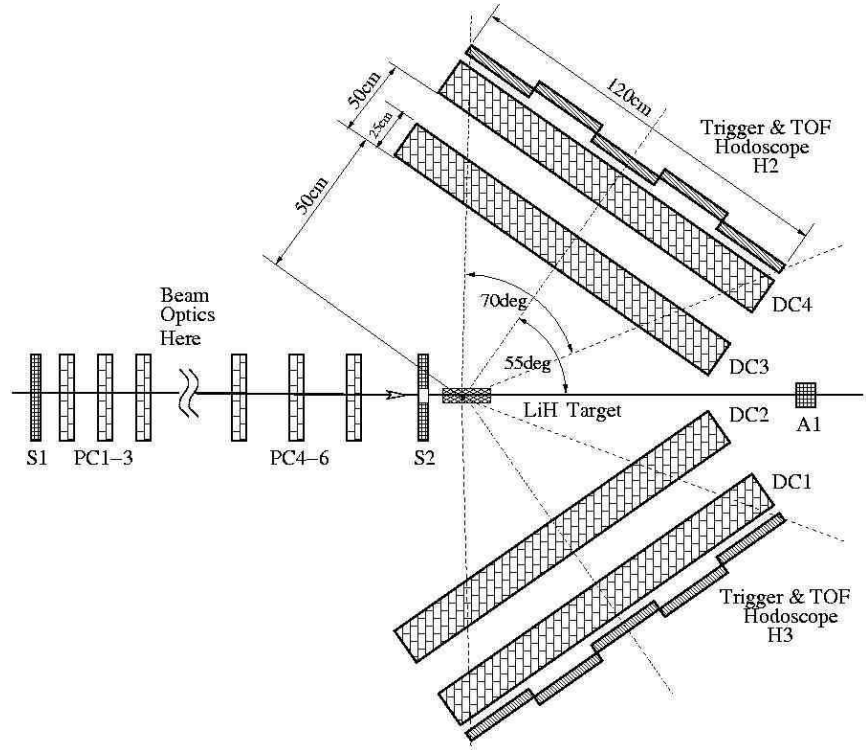

Figure 6. Setup for $N_{\overline{10}}$ search in $\pi^-p$ elastic scattering.

### 1.3.2. Setup geometry

The setup geometry was defined from the requirement of the coverage of c.m. angular interval (40°-120°) including the highly sensitive to $P_{11}$-wave interval (50°-100°) and the interval of low sensitivity, which is essential for the result cross-check. The total azimuthal angle coverage is 2 rad. The angular error is determined by the multiple scattering in the target container, liquid hydrogen and the drift chamber materials. In the proposed experimental scheme the main criteria for the selection of elastic events are the angular correlations of the polar angles of scattered pions and protons and coplanarity. The additional selection criterion is the requirement of the common interaction vertex for the pion and the proton. Therefore it is necessary to use the chambers with minimum amount of light (with large radiation length) matter on the way of the scattered particles to achieve the cleanest experimental result.

In this proposal wire drift chambers with honeycomb cells and modular structure are considered. The sensitive region of each module is 80x120 cm$^2$ in size, the thickness is 25 cm. Each module contains 2 X-planes, 2 Y-planes and 2 stereoplanes. The amount of matter for the module is determined mainly by the external walls of mylar and is equal 0.2–0.3 mm per module. Drift length is 10 mm. One module contains 320 signal wires approximately. The expected coordinate accuracy is less than 0.15 mm. The set of two detectors in each arm provides essential overdetermination of the tracks (in average 5 coordinates in each projection). Such detector structure guarantees high efficiency and stability of the tracking part.

### 1.3.3. Pion Beam and Momentum Measurement

ITEP 10 GeV synchrotron has secondary pion beams which are ideally suitable for the proposed experiment. The energy range, momentum resolution, beam intensity allow to measure the differential cross-sections with a precision an order of magnitude better than in all previous measurements. Currently only ITEP has pion beams of such quality.

The following is the summary of the ITEP 322 beamline features:

- Momentum range 0.8–2.5 GeV/c
- Intensity at the target ($\pi^-$) (2–3)·10$^5$ pions/cycle
- Beam impurities (other particles) <1% total
- Repetition rate 15 cycles/min, up to 1000 ms flattop
- 2-stage achromatic optics
- Spot size in the second focus (10–12) mm FWHM
- Dispersion in the first focus 50 mm/%
- Momentum resolution 0.08% in the interval ±2%
- Transport length 40 m, 3 dipoles, 5 quadrupoles

To measure the pion momentum one needs to define the crossing point of its track with the imaginary focal plane of the first focus. The focal plane is stretched along the beamline for 2.5 m, while allowing the particles to travel such a distance in the air would lead to unwanted increase of the momentum spread due to ionization loss. This is why the set of 3 proportional chambers in the first focus is inserted into only 60 cm air gap in the beamline vacuum pipe and the track is then continued to find the crossing with the focal plane. Such approach requires good coordinate accuracy of the tracking detectors, together with small amount of the material on the way of the particles. Proportional chambers with 1 mm wire pitch suggest a good solution.

The momentum resolution is determined by the size of the internal target image in the first focus of the beamline, which is defined by a number of factors including the shape of the internal target. Yet the momentum resolution ~0.1% can be readily obtained [23], leading to a conservative estimate of the mass resolution as low as $\Delta M_x \approx \Delta p \cdot M_p / M_x = 0{,}56$ MeV at the central pion momentum 1 GeV/c.

The search for a resonance itself does not require the absolute momentum calibration. It is only necessary to maintain the stability of the internal target position and of the dipole magnet fields. But the absolute momentum knowledge is required for the precise determination of the resonance mass, if found. Such calibration can be fulfilled using the process of elastic scattering of protons from an internal polyethylene target with the accuracy ≤0.3%, the error dominated by the knowledge of the proton energy in the accelerator ring.

The field stability of the three dipole magnets must be checked by means of a magnetic field sensor with precision better than 0.1%. Standard instrument based on the nuclear magnetic resonance (NMR) can be used for this purpose. Direct monitoring of the pion beam momentum is also possible by measuring the difference in the time of flight of the negative pions and antiprotons, which are present in the beam at the level of 10$^{-3}$. But the absolute long term stability of such TOF system may appear worse than the required precision [24].

To really use the advantage of high momentum resolution of the beamline one should put as little material on the way of beam particles as possible because of the Landau nature of the ionization losses. Fig. 7 illustrates the simulation results for the momentum distributions of the 1 GeV incident pions at the entrance, in the middle and at the exit of the LiH target, taking into account one 5 mm scintillator counter S1, all mylar windows of the chambers and beampipes and air gaps. The distribution widening is significant, but tolerant within the requirements of this proposal.

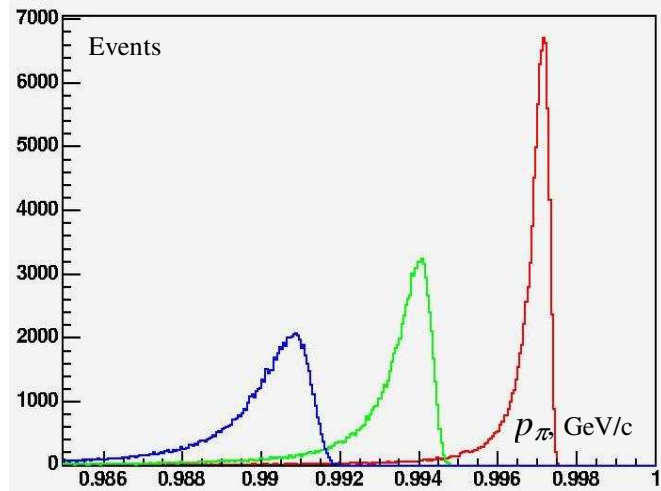

Figure 7. Momentum distribution due to ionization losses of 1 GeV pions in the 25 cm LiH target; red – entrance, green – middle, blue – exit.

### 1.3.4. ToF Technique

The scattered proton identification notably decreases the background under the elastic scattering peak. Such identification can be realized by a simple time of flight system. The time of flight difference of pion and proton between hodoscopes H2 and H3 is ~2.5 ns for c.m. angle ~$75^0$, where the scattered pion and proton are kinematically indistinguishable. Thus the time resolution ~(0.5–1) ns is required, which can be easily achieved.

### *1.4. Setup for Baryon Resonance Study in the Reaction $\pi^-p \to K\Lambda$*

The reaction $\pi^-p \to K\Lambda$ has several properties attractive for the baryon spectroscopy studies in general and for the proposed experiment in particular:
- ❖ Pure isotopic state with isospin ½
- ❖ Sizeable threshold of $K\Lambda$ production facilitating the study of resonances with small spin and large mass
- ❖ High analyzing power of $\Lambda \to \pi^-p$ weak decay with the asymmetry $\alpha=0.642$
- ❖ The threshold of $K\Lambda$ production is lower than the threshold of competitive process with $K^0\Sigma^0$ production
- ❖ Significant fraction of the charged mode (22% of the total reaction cross-section)
- ❖ Large total cross-section (~0.9 mb, [25]) in the considered energy interval

These arguments along with the ones from Sec. 1.1.3 and 1.1.4 makes $\pi^-p \to K\Lambda$ reaction extremely attractive for the search of the cryptoexotic state $N_{\overline{10}}$

### 1.4.1. Specific Properties of π⁻p→KΛ Detection

The setup for $\pi^-p \to K\Lambda$ study can be considered as an extension of the setup for $N_{\overline{10}}$ search in the elastic scattering. The general requirements, listed in Sec. 1.2, are also applicable here. All elements considered in Sec. 1.3 can be reused, the detector placement changed according to the reaction kinematics. More drift chambers are necessary to cover larger acceptance, as well as significant improvements to the TOF system according to more strict requirements of the final proton identification.

The specific kinematic properties of the reaction $\pi^-p \to K\Lambda$ are:

- It is necessary to detect 4 charged particles in the final state ($K^0 \to \pi^+\pi^-$, $\Lambda \to \pi^-p$).
- Most of the particles from $\Lambda$ decays go to the forward hemisphere, but the angles of the pion trajectories can be large relative to the incident pion momentum (see Sec. 1.4.3 and Fig. 9 below).
- The pions from $K^0$ decays have broad angular distribution with long tails covering $90°$ and backward regions
- The identification of a single proton is highly important for the suppression of other inelastic reactions with 4 or more charged particles, as well as for the $\Lambda$ polarization measurement.
- Most of the events have either all 4 particles in the forward direction (~20%), or a proton and 2 pions going forward, the third pion sideward or upward (~80%).

Thus to reach the reasonable acceptance of the setup one needs nearly $4\pi$ hermetic detector system with a large coverage angle for the particles emitted in the forward hemisphere. Unlike the elastic case the TOF hodoscope must be segmented taking into account the multyparticle feature of the reaction, and its time resolution should be improved as will be shown below.

### 1.4.2. Setup Layout

The proposed setup is shown in Fig. 8.

* The outer chambers DC1–DC5 have the size of the sensitive area 80x120 cm², the same as in the elastic setup. The number of chambers is increased from 4 to 5. The chambers cover both sides of the beam, top, bottom and forward directions, forming a parallelepiped open at one side.
* The smaller inner chambers DC6–DC9 cover the angular range in the up/down and left/right directions. The size of the sensitive area is 50x70 cm², the thickness is 10 cm, having only X- and Y- pairs of signal layers.
* The large chamber DC10 with of the size 140x200 cm² is used for forward scattering registration.
* The two-coordinate segmented hodoscope H1,2 is placed at a distance of 2 m from the target and covers the whole sensitive area of DC10, providing the trigger logic and TOF measurement for proton identification. The cell size is 10x10 cm². Such granularity as well as the requirement of good TOF measurement are necessary only in central part of the hodoscope with the size 100x100 cm², because the most of the protons from $\Lambda$-decay go into this region.
* The beam part of the setup including the proportional chambers for the momentum measurement is exactly the same as in the elastic setup and is not shown in the figure.

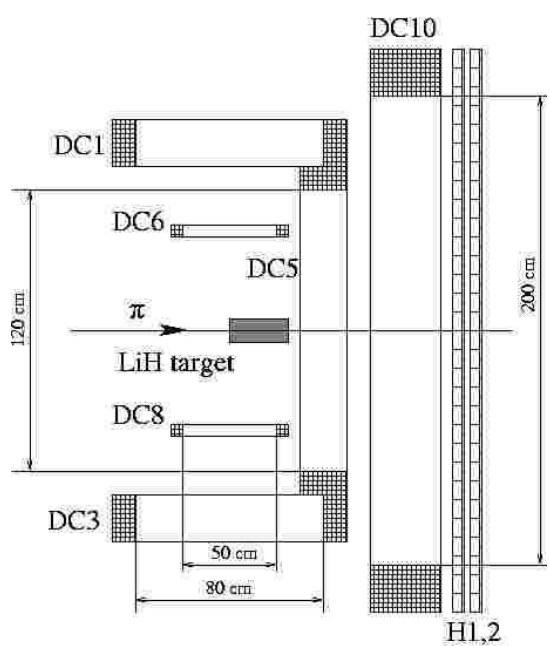

Figure 8. Setup for $N_{\overline{10}}$ search in $\pi^-p \to K\Lambda$ reaction.

### 1.4.3. Some Preliminary Monte-Carlo Simulation Results

The following additional conclusions can be done based on the results of the preliminary MC simulations:

1. The effective registration of $\pi^-p \to K\Lambda$ process in the whole $2\pi$ range of scattering angles is possible without changes to the setup geometry in the energy interval under study. The acceptance varies in the range of (20–30)% of the charged decay mode dependent on the initial energy.

2. There is a significant difference in the kinematics of $K^0\Lambda$ and $K^0\Sigma^0$ production and in the kinematic parameters of secondary particles ($\Sigma^0 \to \Lambda\gamma \to \pi^-p\gamma$), thus allowing the reliable separation of the two reactions in the energy range above $\Sigma^0$ production threshold.

3. Confirmed that the limited spatial and angular distributions of scattered protons (in the forward direction, Fig. 9) and their momentum spread allows the application of TOF methods for their identification. Yet to achieve the reliable separation of the particles by time of flight the resolution of the TOF system should be of the order of (200–300) ns (Fig. 10).

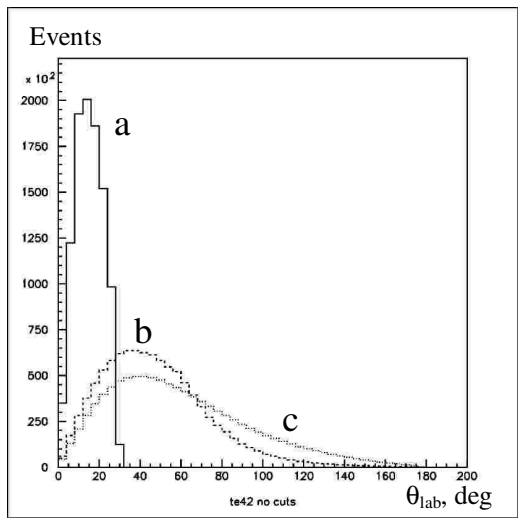
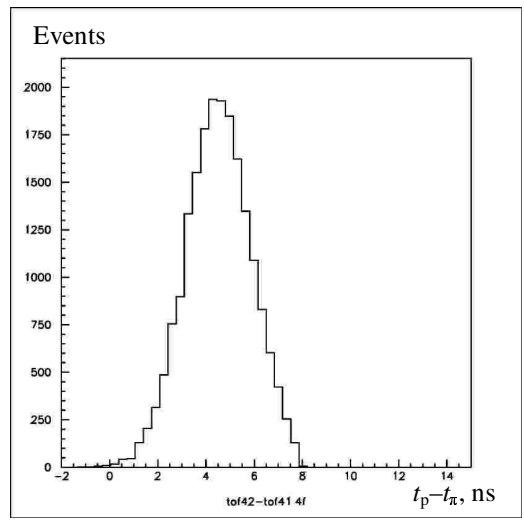

Figure 9. Lab system angular distributions of: a) protons; b) pions from $\Lambda$-decays; c) pions from $K^0$-decays

Figure 10. TOF difference between the proton and the pion from $\Lambda$-decays normalized to 1 m flight distance

4. Confirmed that taking into account the spatial accuracy of the drift chambers (~0.15 mm) and multiple scattering in the materials of the target and the chambers, it's possible to reliably reconstruct and separate both decay vertices of $K^0$ and $\Lambda$, which is absolutely necessary in case of absence of magnetic field for secondary particle momentum measurement.

### *1.5. Counting Rates and Run Times*

#### 1.5.1. Elastic π⁻p→π⁻p scattering

It is proposed to divide the angular range sensitive to $P_{11}$-wave into two subintervals and to reach 0.5% statistical error in each subinterval. In this case the separation of 5% resonant effect would be achieved at 10 standard deviation level. Assuming the bin width in the mass interval (1610–1770) MeV equal to the momentum resolution (0.56 MeV), the total number of momentum-angular intervals in the sensitive range is (160/0.56)x2=570. The necessary total statistics of the elastic scattering events is thus $4 \cdot 10^4$x570=**23x10⁶**.

Using the following assumptions:
- differential cross-section 0.2 mb/sr
- setup acceptance solid angle 1.7 sr
- target thickness 0.071 g/cm³·25 cm=1.78 g/cm²
- beam intensity $50 \cdot 10^3$ s⁻¹ ($200 \cdot 10^3$ per 4 s spill)
- overall efficiency (accelerator + setup) 0,7

the counting rate of the elastic events will be **18.7 s⁻¹** or 75 per an accelerator spill. The trigger rate rough estimate is ~75x5=400 per spill.

The duration of the experiment is: $23 \cdot 10^6/(18.7 \cdot 0.7)=1.75 \cdot 10^6$ s or **20 days**, i.e. 2 runs 10 days each.

In case a narrow resonance is found an additional run of the same 10-day duration is necessary to obtain the statistically valuable detailed angular dependence of the cross-section in order to reliably determine the resonance quantum numbers.

In the angular regions out of the cross-section minimum, where there is no sensitivity to $P_{11}$-wave and the cross-section is significantly higher, the statistical accuracy required for the result cross-check is obtained automaticly.

#### 1.5.2. Reaction π⁻p→KΛ

For the counting rate calculation the total reaction cross-section was taken equal 0.9 mb [25] at 1GeV/c and the acceptance for the charged mode 20% was assumed. The assumptions about the beam and the target parameters are the same as in the elastic scattering case. The calculation results in the counting rate of $125 \cdot 10^3$ events per day. In order to provide the reliable separation of 10% resonant effect in the total reaction cross-section at 10 standard deviation level the statistics of $10^4$x(160/0.56)=**2.9·10⁶** is required, which can be obtained in **23 days** or in 2 runs 12 day each.

## 2. Cost and Time Estimates

Taking into account the growing interest to the study of the "pentaquark" antidecuplet and high activity of different experimental groups to search for this antidecuplet members there is no sense to plan the proposed experiment for time period larger than **3 years**. Such an approach undoubtedly requires the essential concentration of financial investments and manpower.

The following is the proposed schedule of the project:

| | ...→ | Year 1 | Year 2 | Year 3 | Beyond... |
|---|---|---|---|---|---|
| Experiment planning, discussion, project approval, collaboration call, technical design, search for funding | ▨ | | | | |
| Creation of the setup for $N_{\overline{10}}$ search in the elastic scattering, production of the proportional and first 4 modules of the drift chambers, design and production of the electronics for them, 322 beamline resurrection and tests, modification of the LiH target, production of the trigger hodoscopes | | ▨ | | | |
| Data runs for the elastic scattering experiment, start of data processing | | | ▨ | | |
| Production of the additional drift chambers for the setup extension for $\pi^-p \to K\Lambda$ studies, production of the segmented scintillation hodoscope | | | ▨ | | |
| Elastic data processing and publication | | | | ▨ | |
| Data taking for $\pi^-p \to K\Lambda$, start of processing | | | | ▨ | |
| Finalizing of $\pi^-p \to K\Lambda$ data processing, publication | | | | | ▨ |

The total cost of the project was estimated as **40 million rubles**.

## 3. Conclusions

As a conclusion the authors would emphasize several statements concerning this proposal:

◇ The world scientific community reveals big interest to the recently discovered exotic particles ("pentaquarks").

◇ As a result numerous theoretical models got a powerful stimulus for development, requiring an experimental confirmation or denial of their ideas.

◇ Pion beams at ITEP are ideally suitable for the experimental search of the non-strange cryptoexotic member of the pentaquark antidecuplet.

◇ At the moment there are no other pions beams in the world which may offer such possibility.

◇ The experimental conditions are extremely favorable for the suggested experiment: large cross-sections and high sensitivity to the effect, reasonable run time requirements.

◇ If the resonant state is discovered, not only its mass can be determined but also the width (or, at least, an estimate can be made an order of magnitude better than for known antidecuplet members), as well as all the quantum numbers (currently those are not confirmed for any of the "pentaquarks"). It's also possible to extract the branching ratio for the decays to $\pi^-p$ and $K\Lambda$ channels, providing the information about the amount of the strange component in $N_{\overline{10}}$.

◇ If no resonant effect is observed either in $\pi^-p \to \pi^-p$ or in $\pi^-p \to K\Lambda$, such negative result is also very important for the checks and development of theoretical models. In this case rather hard limit will be imposed on the $N_{\overline{10}}$ production cross-section in the two reactions under consideration.

◇ Moreover the data on differential cross-section in both reactions and normal polarization in $\pi^-p \to K\Lambda$ are of great self-importance, it will be readily used by partial wave analyses due to small expected errors and small momentum step of the data points. As a continuation of the experiment, one can imagine taking data in a

- ◇ wider energy and angular range, as well as measurements of $\pi^+p$ elastic scattering, aiming directly to provide excellent quality data for PWA.
- ◇ The discussion of this proposal with theorists showed their high interest to the expected results and the validity of the experimental idea as a whole. Several experimental groups from ITEP and PNPI expressed the readiness to participate in this project by manpower and existing equipment.
- ◇ There are teams at ITEP and PNPI which have a long lasting experience in performing of such type experiments, including their setup, running and data analysis. Currently, both teams in collaboration are finishing the cycle of works on the study of the elastic pion-proton polarization parameters with the SPIN setup at ITEP [26-30]. The two teams may form the base for the new collaboration to carry out this proposed experiment.
- ◇ The proposed experiment is senseless if not completed on a reasonable time scale (3 years). This requires essential concentration of manpower and financial investments. Only united effort of several physics institutions may allow to achieve this goal. Thus

# COLLABORATORS ARE EXTREMELY WELCOME !!!

## Acknowledgments

The authors are grateful to I.I. Strakovsky and Ya.I. Azimov for the very useful discussion of the proposal and for support. The project is partially supported by RFBR grant 05-02-17005.

## Appendix A. Milestones and Plans

The following table summarizes the current status of the experiment and plans for the nearest future, as on April, 1, 2005

| | |
|---|---|
| *Milestones passed* | |
| June 2004 | Proposal submitted at ITEP and PNPI |
| September 2004 | Proposal approved by ITEP scientific council |
| November 2004 | MOU with PNPI was signed, PNPI taking the responsibility for proportional and drift chambers production; ITEP is responsible for the beam/target issues and the chamber electronics |
| December 2004 | 322 beamline at ITEP was resurrected and beam adjustments and tests started, agreement on the LiH target reached between two ITEP groups, technical project of its modification finished |
| March 2005 | Adjustment of beam focusing and tests of the momentum resolution; prototype proportional chamber and electronics production started. Started MC simulations for the elastic setup. |
| April 2005 | LiH target moving to the experiment area started. The project is supported by the Russian Fund for Basic Research, grant 05-02-17005 |
| *Milestones to pass* | |
| June 2005 | Tests of prototypes of proportional and drift chambers and their electronics |
| November 2005 | Liquid neon test run of the target |
| December 2005 | Installation and tests of the whole set of proportional chambers. Tests of the full-scale prototype of drift chamber |
| March 2006 | Liquid hydrogen run of the target |
| June 2006 | Tests of 4 drift chambers for the elastic setup |
| November 2006 | The first data run for the elastic scattering |
| March 2007 | The second data run for the elastic scattering |
| June 2007 | Tests of TOF segmented hodoscope and large drift chamber |
| November 2007 | The first data run for KΛ |
| March 2008 | The second data run for KΛ |